# International Contribution to
# *Nipah Virus* Research 1999-2010

**H. Safahieh[1], S.A. Sanni[1], A.N. Zainab[1,2]**
[1]Department of Library and Information Science,
Faculty of Computer Science & Information Technology, University of Malaya,
50603, Kuala Lumpur, MALAYSIA
[2]Malaysian Citation Centre,
Ministry of Higher Education, Putrajaya, MALAYSIA
e-mail: hsafahieh@yahoo.com; demolasanni@yahoo.com; zainab@um.edu.my

**ABSTRACT**

*This study examines 462 papers on Nipah virus research published from 1999 to 2010, identifying the active authors, institutions and citations received. Data was extracted from SCI-Expanded database, (Web of Science) and analyzed using descriptive figures and tables. The results show the growth of publication is incremental up to 2010 even though the average citations received is decreasing. The ratio of authors to articles is 1330: 426. The active contributing countries are USA (41.0%), Australia (19.3%), Malaysia (16.0%), England (6.5%) and France (5.6%). The productive authors are mainly affiliated to the Centre for Disease Control and Prevention, USA and Commonwealth Scientific and Industrial Research Organization (CSIRO) in Australia and University of Malaya Medical Centre, Malaysia. A total of 10572 citations were received and the ratio of articles to citation is 1: 24.8. Collaboration with the bigger laboratories in USA and Australia is contributive to the sustained growth of published literature and to access diverse expertise.*

## INTRODUCTION

*Nipah virus* is named after Sungai Nipah Village in Perak, Malaysia where the virus was discovered. *Nipah virus encephalitis* outbreak was first reported in September 1998 and documented in 1999 (Anon 1999; Chua et al. 2000). Subsequently, other outbreaks were reported in other Malaysian states by February 1999 and spread to Singapore by March 1999. The main victims were pig farmers or abattoir workers who handled pigs. In Singapore the disease was controlled by ending all importation of pigs from Malaysia (Lam and Chua 2002; Paton et al. 1999). The clinical features presented by victims are fever, headaches, dizziness, vomiting, reduced level of consciousness and brain stem dysfunction. A detailed description of the clinical features was provided by Khean et al. (2000). This virus is classified together with the Hendra virus as a new genus named *Henipa virus* in the subfamily *Paramyxovirinae* (Chong et al. 2006). As a result of this outbreak, over 1 million pigs were culled. It was discovered that the main carrier of the virus are four species of fruit bats. The virus was found in the urine and saliva of infected flying foxes (bats) and pigs consuming food contaminated by these secretions can be infected. This occurs especially when the pig farms are located close to fruit orchards or fruit trees that attracted flying foxes. As the bats are migratory (Eaton and Broder 2006), the alert with regard to the occurrence of this virus spread to South Asian countries. The *Nipah virus* outbreak was severe in Malaysia, with over 200 victims. There have been outbreaks reported in Bangladesh and India (Luby et al. 2006), Thailand (Wacharapluesadee et al. 2005), Cambodia (Reynes et al. 2005), Ghana and Madagascar (Kugler 2004; Chong, Suhailah and Tan 2009). Research on new strategies to inhibit the diseases has spread to other parts of world (Porotto 2011). So far, no study has been conducted to examine the growth and spread of *Nipah virus* research. As the virus was first reported by Malaysian researchers it would be interesting to find out the spread of the research and publication activity on this disease throughout the world. In this study we attempt to analyze published literatures on *Nipah virus* in main stream journals particularly those indexed by the *Web of Science* (*WoS*) for the period 1999-2010. This will assist in tracing the growth trends of *Nipah Virus* research globally.

## OBJECTIVES OF THE STUDY

The objectives of the present study are:
   a. To examine the publication trends on *Nipah Virus* research for the period 1999-2010;
   b. To identify the prolific authors in the field of *Nipah virus*;
   c. To determine authors' productivity and authorship patterns among *Nipah virus* research;
   d. To identify productive institutions researching on *Nipah Virus*;
   e. To determine core journals publishing papers on *Nipah virus* research;
   f. To examine core journals referenced by *Nipah Virus* researchers; and
   g. To identify the citations received by *Nipah Virus* papers.

## RESEARCH METHODOLOGY

This study applied bibliometric approach to gather data on the productivity and research publications of *Nipah Virus*. The research publication on *Nipah Virus* produced from 1999 (initial detection of *Nipah Virus*) to 2010 were searched and retrieved from both the *Web of Science* (*WoS*) and *Scopus* on November 2011. The keywords used to search in the "Topic Search" were "*Nipah Virus*" and articles and review papers were chosen to refine the search. As the number of publications reported in *WoS* was larger, we have subsequently chosen to only use the data set retrieved from *WoS* assuming that the main literature would have been covered. To get the top 100 institutions, the 'institutions' heading was selected, and under the "more options / values" hypertext was used to refine the search. The same procedure applies to obtain the top authors and journals. The retrieved publications were then exported into Microsoft Excel Version 2007 for descriptive analysis. Moreover, Bibliometric toolbox software was employed to calculate the frequency distribution of authors and cited references. This allows us to apply Lokta's law (1926) to the distribution of authors' lists and Bradford law (1948) to the distribution of reference lists respectively.

## RESULTS

### Growth and Distribution of Literature on *Nipah virus*
The total number of articles published in journals indexed by *WoS* was higher (462 papers) than those reported by *Scopus* (413) with a yearly average of 35.5 papers. The growth is incremental and continues to be so in 2010 as indicated by the trendline (Figure 1). Also, the higher degree of success for authors to publish in *SCI* indexed journals may be due to the newness of the discovery and hence any results reported have a higher chance of being published. It may be the case of publisher's publishing anything rather than nothing so that every report on the virus is documented. For subsequent analysis we will use only the larger data set retrieved from *WoS*. The highest number of articles published was recorded in 2010 with 62 articles, followed by 58 articles in year 2009. The results show that research on the subject will continue to grow as indicated by the upward trend line.

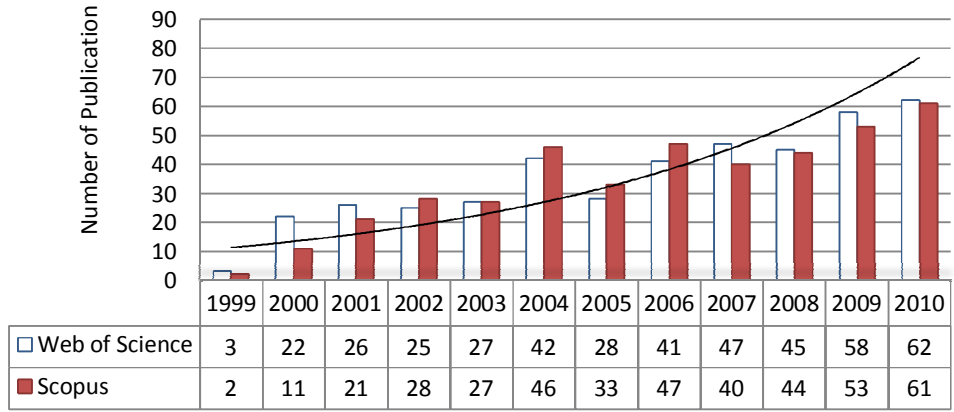

| | 1999 | 2000 | 2001 | 2002 | 2003 | 2004 | 2005 | 2006 | 2007 | 2008 | 2009 | 2010 |
|---|---|---|---|---|---|---|---|---|---|---|---|---|
| Web of Science | 3 | 22 | 26 | 25 | 27 | 42 | 28 | 41 | 47 | 45 | 58 | 62 |
| Scopus | 2 | 11 | 21 | 28 | 27 | 46 | 33 | 47 | 40 | 44 | 53 | 61 |

Figure 1: Year-wise Distribution of Article Publications on *Nipah Virus* (1999-2010)

Figure 2 shows the country distribution of articles on *Nipah Virus* during 1999-2010. The countries ranked in the top five positions for publication contributions are the United States (USA) (189 papers), Australia (89 papers), Malaysia (74 papers), England (30 papers) and France (26 papers). These 5 countries contribute about 88.4% (408) of total articles published. Although *Nipah virus* was initially detected and reported by Malaysian scientists, however, researchers from the USA are currently the most productive contributors to the literatures on this topic.

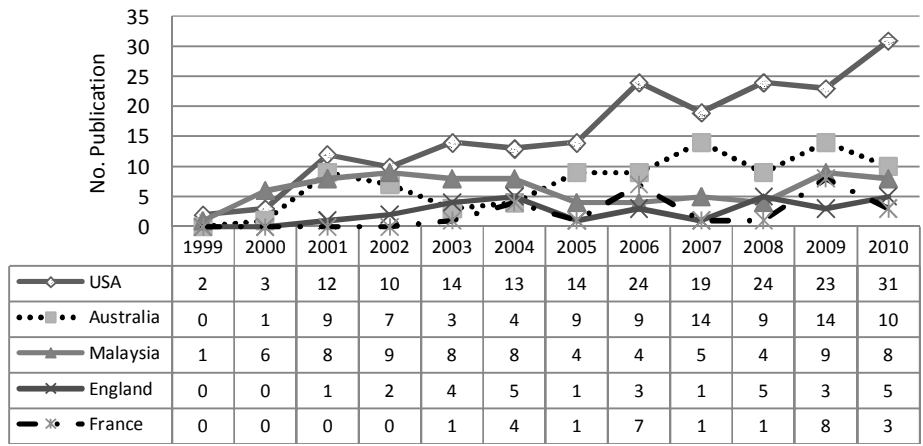

| | 1999 | 2000 | 2001 | 2002 | 2003 | 2004 | 2005 | 2006 | 2007 | 2008 | 2009 | 2010 |
|---|---|---|---|---|---|---|---|---|---|---|---|---|
| USA | 2 | 3 | 12 | 10 | 14 | 13 | 14 | 24 | 19 | 24 | 23 | 31 |
| Australia | 0 | 1 | 9 | 7 | 3 | 4 | 9 | 9 | 14 | 9 | 14 | 10 |
| Malaysia | 1 | 6 | 8 | 9 | 8 | 8 | 4 | 4 | 5 | 4 | 9 | 8 |
| England | 0 | 0 | 1 | 2 | 4 | 5 | 1 | 3 | 1 | 5 | 3 | 5 |
| France | 0 | 0 | 0 | 0 | 1 | 4 | 1 | 7 | 1 | 1 | 8 | 3 |

Figure 2: Year-wise Distribution of Article Publication by Country

## Research Productivity of Authors

A total of 1330 unique authors contributed to the 426 papers in *WoS* during 1999 to 2010. The ratio of the number of authors to articles is 1330: 426 or 1: 0.32. Table 1 shows the most productive authors with their respective author score who have produced at least 7 articles on *Nipah virus*. Wang LF from Australia has the highest author's total scores (7.978) among the top 53 authors. He is also the most prolific author with 51 articles followed by Ksiazek TG (31) Broder CC (28) Eaton BT (26) Rota PA (24). This shows that, Wang LF has produced an average of 5.3 articles per year during the 11-years period, while, Ksiazek TG, Broder CC, Eaton BT, Rota PA have each published an average of 2 articles per year.  Out of the 54 productive authors, 23 authors were from USA, 11 authors from Malaysia, 9 authors from Australia, 4 authors from France, 3 authors from Bangladesh, 2 authors from Germany and 1 author from Canada. This suggest that Malaysia is ranked after Australia in total number of published articles on *Nipah virus*,  but ranked second in  numbers of active authors. This indicates that there are more authors co-authoring papers from Malaysia.

Table 1: Productive Authors with at least Seven Articles from 1999-2010

| No. | Authors | No. of Publication | Country | Author's Total Scores* |
|-----|---------|--------------------|---------|------------------------|
| 1. | Wang L F | 51 | Australia | 7.978 |
| 2. | Ksiazek TG | 31 | USA | 2.975 |
| 3. | Broder CC | 28 | USA | 3.922 |
| 4. | Eaton BT | 26 | Australia | 4.801 |
| 5. | Rota PA | 24 | USA | 3.789 |
| 6. | Crameri G | 23 | Australia | 2.651 |
| 7. | Chua KB | 22 | Malaysia | 4.642 |
| 8. | Rollin PE | 22 | USA | 2.232 |
| 9. | Lee Be | 21 | USA | 3.364 |
| 10. | Lam SK | 19 | Malaysia | 3.555 |
| 11. | Daszak P | 17 | USA | 3.137 |
| 12. | Tan CT | 18 | Malaysia | 3.446 |
| 13. | Wong KT | 17 | Malaysia | 5.013 |
| 14. | Bellini WJ | 16 | USA | 2.373 |
| 15. | Bossart KN | 16 | USA | 1.762 |
| 16. | Field HE | 15 | Australia | 3.599 |
| 17. | Hossain MJ | 16 | Bangladesh | 1.723 |
| 18. | Aguilar HC | 15 | USA | 1.581 |
| 19. | Mungall BA | 14 | USA | 2.372 |
| 20. | Dutch RE | 13 | USA | 5.259 |
| 21. | Gurley, ES | 12 | Bangladesh | 1.315 |
| 22. | Halpin K | 12 | Australia | 2.144 |
| 23. | Luby SP | 12 | USA | 1.321 |
| 24. | Maisner A | 12 | Germany | 3.232 |
| 25. | Middleton D | 12 | Australia | 1.56 |
| 26. | Tan, WS | 11 | Malaysia | 2.533 |
| 27. | Wild, TF | 11 | France | 2.114 |
| 28. | Yu, M | 11 | Australia | 1.231 |
| 29. | Comer, J. A. | 10 | USA | 0.761 |
| 30. | Diederich,S | 10 | Germany | 2.566 |
| 31. | Goh, KJ | 10 | Malaysia | 1.165 |
| 32. | Harcourt, BH | 10 | USA | 1.405 |
| 33. | Horvath, CM | 10 | USA | 4.824 |
| 34. | Zaki, SR | 10 | USA | 1.155 |
| 35. | Chong, HT | 9 | Malaysia | 1.671 |
| 36. | Czub, M | 9 | Canada | 1.399 |
| 37. | Dimitrov, DS | 8 | USA | 0.838 |
| 38. | Guillaume, V | 8 | France | 0.943 |
| 39. | Lo, MK | 8 | USA | 1.564 |
| 40. | Moscona, A | 8 | USA | 0.974 |
| 41. | Porotto, M | 8 | USA | 0.974 |
| 42. | Tamin, A | 8 | USA | 0.955 |
| 43. | Basler, CF | 7 | USA | 1.761 |
| 44. | Breiman, RF | 7 | Bangladesh | 0.454 |
| 45. | Buckland R | 7 | France | 1.009 |
| 46. | Chang , LY | 7 | Malaysia | 1.571 |
| 47. | Epstein, JH | 7 | Malaysia | 0.764 |
| 48. | Georges-Courbot, MC | 7 | France | 0.723 |
| 49. | Hassan, SS | 7 | Malaysia | 1.196 |
| 50. | McEachern, JA | 7 | Australia | 0.676 |
| 51. | Michalski, WP | 7 | Australia | 1.104 |
| 52. | Shaw, ML | 7 | USA | 2.071 |
| 53. | Wolf, MC | 7 | USA | 0.904 |
| 54. | Yusoff, K | 7 | Malaysia | 1.65 |
| | Rest of 1276 authors | Less than 7 | | |

## Authorship Pattern

Previous bibliometric studies show that normally research in the sciences and medical sciences are commonly carried out by group of researchers rather than by a single researcher (Abrizah and Wee 2011; Zainal and Zainab 2011). A study by Melin (2000) found that through collaboration, researchers may increase their knowledge, improve the quality of research, establish contacts and networks for future research, generate new ideas and become more productive in terms of publishing papers. In this study we examine the authorship pattern in the *Nipah virus* researches. The result shows a bias towards mega-authorship (244 articles, 57.3%), where in this context are papers authored by five or more authors. The rest are 3-4 authors (95, 22.3%), 2 authors (41, 9.6%) and single authored works (46, 10.8%).

## Frequency Distribution of Authors Productivity (Lokta's Law)

We applied Loktas's law to find out about the frequency distribution of scientific productivity. The results show that one author made 51 contributions to *"Nipah Virus"* research between 1999 - 2010, while another made 31contributions, and so on (Table 2).

Table 2: Author's Productivity Pattern Observed Compared with Expected (Lokta)

| No of Contribution (n) | No of authors | Predicted no of authors | Differences |
|---|---|---|---|
| 51 | 1 | 0.37 | 0.63 |
| 31 | 1 | 1 | 0 |
| 28 | 1 | 1.22 | 0.22 |
| 26 | 1 | 1.42 | 0.42 |
| 24 | 1 | 1.66 | 0.66 |
| 23 | 1 | 1.81 | 0.81 |
| 22 | 2 | 1.98 | 0.02 |
| 19 | 1 | 2.65 | 1.65 |
| 18 | 1 | 2.95 | 1.95 |
| 17 | 3 | 3.31 | 0.31 |
| 16 | 2 | 3.74 | 1.74 |
| 15 | 2 | 4.25 | 2.25 |
| 14 | 1 | 4.88 | 3.88 |
| 13 | 1 | 5.66 | 4.66 |
| 12 | 6 | 6.65 | 0.65 |
| 11 | 3 | 7.91 | 4.91 |
| 10 | 6 | 9.57 | 3.57 |
| 9 | 2 | 11.81 | 9.81 |
| 8 | 6 | 14.95 | 8.95 |
| 7 | 12 | 19.53 | 7.53 |
| 6 | 15 | 26.58 | 11.58 |
| 5 | 21 | 38.28 | 17.28 |
| 4 | 32 | 59.81 | 27.81 |
| 3 | 71 | 106.33 | 35.33 |
| 2 | 180 | 239.25 | 59.25 |
| 1 | 957 | 798 | 159 |
| | 1330 | | |

The majority, 957 authors are one time contributors. Thus, by applying Lokta's law, we seek to examine whether "the number (of authors) making n contributions is about $1 / n^c$ of those making one contribution, where $c$ nearly always equals two ($c \approx 2$) ; and the proportion of all contributors, that makes a single contribution, is about 60 per cent." (Lotka 1926). Table 3 presented the differences between the observed numbers of authors with frequency of occurrence against the Lokta's assumed numbers of authors. We could see that there is a slight difference between our results and Lotka's findings. Lotka found single contributors to be about 60 percent in his own examination, whereas we found the proportion that makes a single contribution is 71.9%, and few authors contributed more than one article. The result is similar to Sanni and Zainab (2010) who reported that the percentage of authors that make just one contribution to the *Medical Journal of Malaysia* (from 2004 -2008) is 63.4%.  This is also parallel with the findings by Chung and Cox (1990) for contributors to finance literatures.

### Research Productivity by Institutional Affiliation

In order to examine the most productive institutions on *Nipah virus* researches we searched the first top 100 institutions option of the WoS database. Table 3 represents the list of world-wide productive institutions (sorted by record count) which have published at least 5 articles on *Nipah virus* during the eleven years of study.

Table 3: Productive Institutions with at least 5 Articles during 1999-2000

| No. | institutions | No. of Publication | Country |
|---|---|---|---|
| 1. | Center for Disease Control and Prevention | 55 (12.94%) | USA |
| 2. | University of Malaya | 49 (11.53%) | Malaysia |
| 3. | Commonwealth Scientific and Industrial Research Organization (CSIRO) | 45 (10.59%) | Australia |
| 4. | Uniformed Services University of the Health Sciences | 28(6.59%) | USA |
| 5. | CSIRO Livestock Industries | 27(6.35%) | Australia |
| 6. | University of California, Los Angeles | 19(4.47%) | USA |
| 7. | University of Kentucky | 13 (3.06%) | USA |
| 8. | University of Marburg | 13 (3.06%) | Germany |
| 9. | Universiti Putra Malaysia | 13 (3.06%) | Malaysia |
| 10. | Veterinary Research Institute | 13 (3.06%) | Malaysia |
| 11. | Consortium for Conservation Medicine | 10(2.35%) | USA |
| 12. | University of Queensland | 10 (2.35%) | Australia |
| 13. | Cornell University | 9 (2.12%) | USA |
| 14. | Emory University | 9 (2.12%) | USA |
| 15. | National Institute for Health and Medical Research | 9 (2.12%) | France |
| 16. | Ministry of Health | 9 (2.12%) | Singapore |
| 17. | Mount Sinai School of Medicine | 9 (2.12%) | USA |
| 18. | Northwestern University | 9 (2.12%) | USA |
| 19. | National Institute of Animal Health (NIAH) | 8 (1.88%) | Japan |
| 20. | Singapore General Hospital | 8 (1.88%) | Singapore |
| 21. | University of Lyon | 8 (1.88%) | France |
| 22. | Australian Animal Health Lab | 7 (1.65%) | Australia |
| 23. | Department of Primary Industries and Fisheries | 7 (1.65%) | Australia |
| 24. | Iowa State University | 7 (1.65%) | USA |
| 25. | National Institute of Allergy and Infectious Diseases | 7 (1.65%) | Singapore |
| 26. | National Institute of Allergy and Infectious Diseases | 7 (1.65%) | USA |
| 27. | University of Manitoba | 7 (1.65%) | Canada |
| 28. | Canadian Food Inspect Agency | 6 (1.41%) | Canada |
| 29. | Institute Pasteur | 6 (1.41%) | France |
| 30. | National Cancer Institute | 6 (1.41%) | USA |
| 31. | Queensland Department of Primary Industries | 6 (1.41%) | Australia |
| 32. | Tan Tock Seng Hospital | 6 (1.41%) | Singapore |
| 33. | University of Georgia | 6 (1.41%) | USA |
| 34. | University of Penn | 6 (1.41%) | USA |
| 35. | University of Tokyo | 6 (1.41%) | Japan |
| 36. | Australian Bio Security Coop Research Centre for Emerging | 5 (1.18%) | Australia |
| 37. | CUNY Mount Sinai School Of Medicine | 5 (1.18%) | USA |
| 38. | Department of Veterinary Services | 5 (1.18%) | Malaysia |
| 39. | International Centre for Diarrhoeal Disease Research | 5 (1.18%) | Bangladesh |
| 40. | Institute of Epidemiology Disease Control and Research | 5 (1.18%) | Bangladesh |
| 41. | University of California Davis | 5 (1.18%) | USA |
| 42. | University of Oxford | 5 (1.18%) | England |
| 43. | University of Texas | 5 (1.18%) | USA |
| 44. | WHO | 5 (1.18%) | Bangladesh |

The results show that a significant number (149, 35%) of articles were produced individually or collaborative by only three institutions including Centers for Disease Control and Prevention, USA (55, 13%), University of Malaya, Malaysia (49, 11.53%) and Commonwealth Scientific and Industrial Research Organization, Australia (45, 10. 59%). Out of 44 productive institutions, 17 institutes are from the USA, 7 from Australia, 4 each from Malaysia and Singapore, 3 each from Bangladesh and France, 2 from Japan, and 1 each from Germany and England. This result is expected because researchers from the University of Malaya were the ones who first discovered the outbreak. Soon after the virus was detected, the Center for Disease Control and Prevention in Atlanta, USA was consulted to share information and details about the new epidemic. Thereafter professionals from Singapore and Australia were meeting with their Malaysian counterparts to combat the disease head-on (Ling 1999; Chua et al. 2000; Wong et al. 2002).

### The Core Journals in *Nipah Virus* Research

Examining the literature on *Nipah virus* during the period of 1999- 2010, revealed that a total of 426 articles were published in 157 journals. The journals, which published at least 3 articles on *Nipah Virus* with their respective impact factor (JCR 2010) is listed in Table 4. A significant number of articles (107, 22.8%) were published in only 3 journals; *Journal of Virology* (57, 13.4%), *Virology* (27, 6.3%), and *Emerging Infectious Diseases* (23, 5.4%). The rest of the papers were published in 154 Journals (319, 77%). Out of 32 journals which contain at least 3 articles on *Nipah virus*, the highest number (16) of journals are published in the USA, followed by England (9), Netherland (3), France (2), Malaysia and Austria (1) each. This indicates that most of the articles are published in foreign mainstream journals on virology research which are, mainly published in the USA.

Table 4: Journal Titles Publishing at Least 3 Articles during 1999-2010

| No. | Source Title | No. of Article | JCR 2010 Impact Factor | Country |
|---|---|---|---|---|
| 1. | *Journal of virology* | 57 | 5.189 | USA |
| 2. | *Virology* | 27 | 3.305 | USA |
| 3. | *Emerging infectious diseases* | 23 | 6.859 | USA |
| 4. | *Journal of general virology* | 13 | 3.568 | England |
| 5. | *Journal of virological methods* | 13 | 2.139 | Netherlands |
| 6. | *Virology journal* | 11 | 2.546 | England |
| 7. | *Archives of virology* | 10 | 2.209 | Austria |
| 8. | *Neurology Asia* | 10 | 0.531 | Malaysia |
| 9. | *Microbes and infection* | 9 | 2.726 | France |
| 10. | *Virus research* | 9 | 2.905 | Netherlands |
| 11. | *Plos pathogens* | 7 | 9.079 | USA |
| 12. | *Proc. of the Natl. academy of sciences of the United States of America* | 7 | 9.771 | USA |
| 13. | *Revue scientifique et technique de l office international des epizooties* | 7 | 1.609 | France |
| 14. | *Clinical infectious diseases* | 6 | 8.186 | USA |
| 15. | *Journal of infectious diseases* | 6 | 6.288 | USA |
| 16. | *Plos one* | 6 | 4.411 | USA |
| 17. | *Annals of neurology* | 5 | 10.746 | USA |
| 18. | *Annals of the new York academy of sciences* | 5 | 2.847 | USA |
| 19. | *Current topics in microbiology and immunology* | 5 | 4.121 | USA |
| 20. | *Comparative immunology microbiology & infectious diseases* | 4 | 3.605 | England |
| 21. | *Journal of comparative pathology* | 4 | 1.529 | England |
| 22. | *Journal of medical virology* | 4 | 2.895 | USA |
| 23. | *Journal of neurovirology* | 4 | 2.243 | USA |
| 24. | *Lancet* | 4 | 33.633 | England |
| 25. | *Australian veterinary journal* | 3 | 1.006 | England |
| 26. | *Current opinion in neurology* | 3 | 4.121 | USA |
| 27. | *Ecohealth* | 3 | 1.640 | USA |
| 28. | *Journal of biological chemistry* | 3 | 5.328 | USA |
| 29. | *Journal of clinical virology* | 3 | 4.023 | Netherlands |
| 30. | *Journal of neurology neurosurgery and psychiatry* | 3 | 4.791 | England |
| 31. | *Nature* | 3 | 36.104 | England |
| 32. | *Vaccine* | 3 | 3.572 | England |
| | Total | 280 | | |

### Bradford's Zonal Analysis of Highly Referenced Journals

The study also seeks to identify the most referenced journal titles in *Nipah Virus* research. According to Bradford (1948): "if scientific journals are arranged in order of decreasing productivity," we should be able to identify core journals devoted to the subject and several groups or zones containing the same number of articles as the nucleus and the number of periodicals in the nucleus and the succeeding zones will be as 1: b: b² (Glanzel 2003). To determine core journals in the literature, we applied *Bradford's Law of Scattering* to the reference list of the articles and created 3 zones (Table 5), each producing approximately one third of the total reference cited.

Table 5: Core Journal Referenced with Zone

| Zones | Journal Title | Frequency of citation |
|---|---|---|
| **Zone 1** | | |
| 1 | *J VIROL* | 2231 |
| 2 | *VIROLOGY* | 1215 |
| 3 | *EMERG INFECT DIS* | 1210 |
| 4 | *SCIENCE* | 805 |
| 5 | *LANCET* | 608 |
| 6 | *J GEN VIROL* | 553 |
| **6 journal count** | | **6622** |
| **Zone 2** | | |
| 7 | *MICROBES INFECT* | 479 |
| 8 | *P NATL ACAD SCI USA* | 464 |
| 9 | *NATURE* | 377 |
| 10 | *NEW ENGL J MED* | 310 |
| 11 | *J INFECT DIS* | 292 |
| 12 | *AUST VET J* | 225 |
| 13 | *J BIOL CHEM* | 219 |
| 14 | *VIRUS RES* | 206 |
| 15 | *ANN NEUROL* | 172 |
| 16 | *ARCH VIROL* | 169 |
| 17 | *CLIN INFECT DIS* | 169 |
| 18 | *AM J TROP MED HYG* | 164 |
| 19 | *MMWR-MORBID MORTAL W* | 155 |
| 20 | *J COMP PATHOL* | 154 |
| 21 | *J VIROL METHODS* | 137 |
| 22 | *REV SCI TECH OIE* | 132 |
| 23 | *J CLIN MICROBIOL* | 129 |
| 24 | *MED J AUSTRALIA* | 121 |
| 25 | *FIELDS VIROLOGY* | 119 |
| 26 | *EMBO J* | 116 |
| 27 | *J CLIN VIROL* | 112 |
| 28 | *AM J PATHOL* | 108 |
| 29 | *HLTH SCI B* | 102 |
| 30 | *CELL* | 89 |
| 31 | *VACCINE* | 88 |
| 32 | *NEUROL J SE ASIA* | 75 |
| 33 | *NAT REV MICROBIOL* | 74 |
| 34 | *CURR TOP MICROBIOL* | 73 |
| 35 | *VET MICROBIOL* | 71 |
| 36 | *J CELL BIOL* | 70 |
| 37 | *J IMMUNOL* | 68 |
| 38 | *J MED VIROL* | 63 |
| 39 | *ANTIVIR RES* | 62 |
| 40 | *VIROL J* | 59 |
| 41 | *AM J NEURORADIOL* | 57 |
| 42 | *PLOS PATHOG* | 54 |
| 43 | *VET REC* | 54 |
| 44 | *VET PATHOL* | 52 |
| 45 | *BIOCHEM BIOPH RES CO* | 51 |
| 46 | *CLIN MICROBIOL REV* | 50 |
| 47 | *NUCLEIC ACIDS RES* | 50 |
| 48 | *J MOL BIOL* | 49 |
| 49 | *J NEUROL NEUROSUR PS* | 49 |
| 50 | *ANTIMICROB AGENTS CH* | 47 |
| 51 | *GENE* | 46 |
| 52 | *NAT MED* | 46 |
| 53 | *JAMA-J AM MED ASSOC* | 43 |
| 54 | *PHILOS T ROY SOC B* | 41 |
| 55 | *SE ASIAN J TROP MED* | 41 |
| 56 | *EPIDEMIOL INFECT* | 40 |
| 57 | *ANNU REV BIOCHEM* | 39 |
| 58 | *AVIAN DIS* | 39 |
| **52 Journal counts** | | **6271** |
| **Zone 3** | | |
| **1862 journal counts** | | **5703** |

We find only 6 journals make the zone 1 list. This means that research publication is concentrated in few high impact journals. The research area have not span more than 11 years and this may be the explanation as to the reason for the literatures been highly published in selected few journals. The research in this field is expected to diffuse across other medical related fields and journals in the future. Hence, we found 6 journals (6622 references) in (zone 1), 52 journals (6271 references) in (Zone 2), and 1862 journals (5703 references) in (Zone 3). The core journal titles which are in the nucleus zone are journals that have citations >= 553. This result is in line with Bradford's Law of Scattering, due to the fact that the journal titles showed a wide dispersion among very small core journals, with less than one percent of the journals accounting for one-thirds of all the cited references.

### Highly Referenced Papers

We identify and distinguish the highly referenced papers on *Nipah Virus*. From Table 5 we identify the top most cited journal titles are *"Journal of virology" "Virology" "Emerging Infectious Diseases" "Science" "Lancet' and "Journal of General Virology"*. Accordingly, the most referenced article is titled: *"Nipah virus: A recently emergent deadly paramyxovirus"* was published in the journal: *"Science"* in year 2000. The paper was co-authored by authors affiliated to Malaysia, USA, Australia, and Singapore. As observed, the most referenced papers were published in the top ranked journals and most of the cited papers are co-authored papers. Most of these highly referenced papers are those wherein the outbreak of the *"Nipah Virus"* was first reported. While others are those in which other similar virus were studied (Table 6).

Table 6: Highly Referenced Papers

|   | Paper Title | Year Published | Times Referenced | Authors Affiliation |
|---|---|---|---|---|
| 1 | Nipah Virus: a Recently Emergent Deadly Paramyxovirus | 1999 | 240 | MAL, USA, AUS & SING |
| 2 | Fatal Encephalitis Due to Nipah Virus Among Pig-Farmers in Malaysia | 2000 | 165 | MAL & USA |
| 3 | A Morbillivirus that Caused Fatal Disease in Horses and Humans | 1995 | 154 | AUSTRALIA |
| 4 | Isolation of Nipah Virus From Malaysian Island Flying-Foxes | 2002 | 128 | MALAYSIA |
| 5 | Molecular Characterization of Nipah Virus, a Newly Emergent Paramyxoviru | 2000 | 113 | USA |

### Citations Received by *Nipah Virus* Articles

Citation analysis is the frequency with which papers published in a field are cited by other papers. It explains the citation count for a journal, an article, a field, or a country's publications (Chiu and Ho 2005). We found that *Nipah Virus* papers accumulated a total of 10,572 citations over the 11-year period (Table 7). The year 1999 recorded the highest rate (140.67 citations) of "average citation per article". This is expected, since this was the year the virus was discovered and subsequent papers are likely to make reference to the first papers reporting the discovery of the virus.

Furthermore, as expected, old papers are more likely to have received more citations than later papers due to sufficient year lag to allow for accumulation of citations. Nonetheless, papers of recent years are also heavily cited and most of the citations (73.76%) are from journal articles (73.76%), followed by reviews (16.03%), and conference proceedings (4.98%). Very few citations were recorded from books (0.93%) and no citations from thesis and dissertations. We found while authors researching on *"Nipah Virus"* referenced mainly journal articles (97.64%), their papers are receiving widespread citations from journal articles to biographies. Table 8 illustrates the country affiliation of authors citing *"Nipah*

*Virus"* research papers. 93 countries were observed from 11 regions of the world. Among these authors, 36.4% are affiliated to the USA followed by Australia (6.03%), England (5.7%), Germany (5.63%), China (5.61%), France (5.13%), Canada (4.07%), Japan (3.97%), Malaysia (2.28%), and Netherlands (2.03%).

Table 7: Details of Citations Received by *Nipah Virus* Articles

| Year | No of Articles Published | No of Citations Received | No without self citation | No of Citing Articles | Average citation per Paper | h-index |
|------|------|------|------|------|------|------|
| 1999 | 3 | 422 | 422 | 293 | 140.67 | 3 |
| 2000 | 22 | 1162 | 1156 | 599 | 52.82 | 13 |
| 2001 | 26 | 1148 | 1133 | 678 | 44.15 | 17 |
| 2002 | 25 | 1018 | 1005 | 691 | 40.72 | 19 |
| 2003 | 27 | 1019 | 1016 | 862 | 37.74 | 17 |
| 2004 | 42 | 1283 | 1269 | 954 | 30.55 | 20 |
| 2005 | 28 | 1268 | 1250 | 841 | 45.29 | 18 |
| 2006 | 41 | 1149 | 1127 | 696 | 28.02 | 20 |
| 2007 | 47 | 742 | 729 | 503 | 15.79 | 18 |
| 2008 | 45 | 790 | 782 | 635 | 17.56 | 14 |
| 2009 | 58 | 381 | 356 | 231 | 6.57 | 11 |
| 2010 | 62 | 190 | 175 | 138 | 3.06 | 6 |
| **Total** | **426** | **10572** | **10420** | **7121** | **24.88** | |

Table 8: Country Affiliations of Authors Citing *Nipah Virus* Papers

| AFRICA | | CARRIBEAN | | EUROPE | | NORTH AMERICA | |
|------|------|------|------|------|------|------|------|
| SOUTH AFRICA | 18 | TRINID TOBAGO | 5 | ENGLAND | 275 | USA | 1751 |
| KENYA | 12 | BARBADOS | 1 | GERMANY | 271 | CANADA | 196 |
| GABON | 6 | CUBA | 1 | FRANCE | 247 | MEXICO | 18 |
| MADAGASCAR | 5 | GUADELOUPE | 1 | NETHERLANDS | 98 | | **1965** |
| UGANDA | 5 | | **8** | SWITZERLAND | 81 | | |
| CAMEROON | 4 | | | SCOTLAND | 72 | SOUTH AMERICA | |
| GHANA | 4 | CENTRAL AMERICA | | SPAIN | 63 | BRAZIL | 35 |
| NIGERIA | 4 | COSTA RICA | 4 | ITALY | 59 | CHILE | 13 |
| REUNION | 4 | PANAMA | 4 | SWEDEN | 40 | ARGENTINA | 10 |
| TANZANIA | 4 | | **8** | BELGIUM | 35 | COLOMBIA | 7 |
| EGYPT | 3 | | | AUSTRIA | 24 | PERU | 5 |
| BOTSWANA | 2 | EAST ASIA | | DENMARK | 21 | VENEZUELA | 4 |
| CONGO | 2 | CHINA | 270 | PORTUGAL | 16 | ECUADOR | 3 |
| ETHIOPIA | 2 | JAPAN | 191 | POLAND | 14 | FRENCH GUINA | 2 |
| ZIMBABWE | 2 | TAIWAN | 47 | LITHUANIA | 13 | URUGUAY | 1 |
| MOROCCO | 1 | SOUTH KORA | 33 | NORTH IRELAND | 12 | | **80** |
| SENEGAL | 1 | | **541** | NORWAY | 12 | | |
| TUNISIA | 1 | | | FINLAND | 11 | SOUTH ASIA | |
| | **80** | MIDDLE EAST | | HUNGARY | 10 | INDIA | 70 |
| | | ISRAEL | 23 | IRELAND | 10 | BANGLADESH | 22 |
| SOUTHEAST ASIA | | TURKEY | 13 | RUSSIA | 9 | PAKISTAN | 5 |
| MALAYSIA | 110 | SYRIA | 2 | GREECE | 7 | SRI LANKA | 3 |
| SINGAPORE | 61 | JORDAN | 1 | CZECH REPUBLIC | 5 | | **100** |
| THAILAND | 32 | KUWAIT | 1 | SLOVAKIA | 4 | | |
| VIETNAM | 13 | SAUDI ARABIA | 1 | SLOVENIA | 4 | AUSTRALASIA | |
| CAMBODIA | 7 | | **41** | BULGARIA | 3 | AUSTRALIA | 290 |
| INDONESIA | 7 | | | LUXEMBOURG | 3 | NEW ZEALAND | 27 |
| PHILIPPINES | 6 | | | SERBIA | 2 | PAPUA N GUINEA | 1 |
| NEPAL | 3 | | | WALES | 2 | | **318** |
| LAOS | 2 | | | CROATIA | 1 | | |
| | **241** | | | ICELAND | 1 | | |
| | | | | | **1425** | | |

Moreover, the study observed that *"Nipah Virus"* received citations from top high ranked journals in virology, such as *Journal of Virology* with 391 counts followed by *Virology* (133 counts), *Emerging Infectious Diseases* (107 counts), *Journal of General Virology* (74 counts), *Virus Research* (69), and *Plos One* (47 counts) (Table 9).

Table 9: Top Journals Citing *"Nipah Virus"* Articles

|  | Journal Titles | Frequency |
|---|---|---|
| 1 | JOURNAL OF VIROLOGY | 391 |
| 2 | VIROLOGY | 133 |
| 3 | EMERGING INFECTIOUS DISEASES | 107 |
| 4 | JOURNAL OF GENERAL VIROLOGY | 74 |
| 5 | VIRUS RESEARCH | 69 |
| 6 | PLOS ONE | 47 |
| 7 | VACCINE | 43 |
| 8 | PROCEEDINGS OF THE NAT. ACADEMY OF SCIENCES OF THE USA | 40 |
| 9 | ARCHIVES OF VIROLOGY | 38 |
| 10 | JOURNAL OF VIROLOGICAL METHODS | 33 |
| 11 | PLOS PATHOGENS | 31 |
| 12 | ECOHEALTH | 31 |
| 13 | REVUE SCIENTIFIQUE ET TECHNIQUE DE L OFFICE INTERNATIONAL DES EPIZOOTIES | 31 |
| 14 | VIROLOGY JOURNAL | 30 |
| 15 | JOURNAL OF BIOLOGICAL CHEMISTRY | 29 |
| 16 | JOURNAL OF IMMUNOLOGY | 29 |
| 17 | JOURNAL OF INFECTIOUS DISEASES | 23 |
| 18 | ANTIVIRAL RESEARCH | 23 |
| 19 | XENOTRANSPLANTATION | 21 |
| 20 | JOURNAL OF CLINICAL VIROLOGY | 20 |
| 21 | SCIENCE | 20 |
| 22 | NATURE REVIEWS MICROBIOLOGY | 19 |
| 23 | VIRUSES-BASEL | 19 |
| 24 | FUTURE VIROLOGY | 18 |
| 25 | MICROBES AND INFECTION | 18 |
| 26 | AMERICAN JOURNAL OF TROPICAL MEDICINE AND HYGIENE | 18 |
| 27 | CLINICAL INFECTIOUS DISEASES | 16 |
| 28 | VETERINARY MICROBIOLOGY | 15 |
| 29 | JOURNAL OF MEDICAL VIROLOGY | 15 |
| 30 | PROCEEDINGS OF THE ROYAL SOCIETY B-BIOLOGICAL SCIENCES | 14 |

## CONCLUSION

Our findings indicate that published research on *Nipah Virus* is still on the increase and especially so from 2009 onwards. Also, the coverage on this subject in universal citation indexes is almost similar with a higher number of articles recorded in *WoS*. The productive authors, who produced at least 7 articles mainly comes from three countries, the USA, Australia and Malaysia. In fact most of the papers published and cited are produced by collaborating authors from the three countries, especially between researchers from the University of Malaya Medical Centre, the Centre for Disease Control and Prevention, USA and Commonwealth Scientific and Industrial Research Organization (CSIRO) in Australia. This in turn is reflected by the mega-authorship pattern of published papers. The journals the researchers have chosen to publish in are mainly those covering virology issues, and are important journals in this field. From the list of journals publishing articles on *Nipah Virus* only one journal is published in Malaysia. *Neurology Asia* contributed 10 papers and is published by the University of Malaya Medical Centre for the Asean Neurological Association and is indexed by *WoS* since 1996. The citations received by research on *Nipah Virus* is large. The ration of paper to citation is 1 : 24.8. Average citation per paper was 40 and above in the first five years of its discovery and this tapers down gradually even though the number of articles published is increasing. This may indicate that the field needs a new focus or it may indicate the maturity level of this research. Virology science is often grounded on collaborative efforts as it attempts to map unknown territory. However, the field remain "small scale" or may be categorized as "mezzo science" (Vermeulen et al. 2010), which, even though it involves complex coordination and involvement of diverse expertise, it is focused on a specific objectives or data and studied on a smaller scale.

In general, the findings show that *"Nipah Virus"* research is a subject that has generated global concern and this have opens floodgates of research conducted with the goal of battling and controlling the epidemic. The rapid growth of research papers on *"Nipah Virus"* within a short period, explains the characteristics of how knowledge flows in scholarly communication. The research was discovered in a local area (Malaysia) and reported through the main research communication channel (journals). Researchers from other part of the world were able to find out about this new discovery and immediately engage and contribute to the research and discussion. Some years after, result is now showing that the most productive *"users"* of the research reports are those from the most developed countries (USA, Australia, England, Germany, China, France, Canada, Japan, Malaysia and Netherlands). Malaysia is probably in the top ten because the virus was discovered in Malaysia. Our results further highlights the fact that the most developed countries of the world are still the most active in researching on the *Nipah Virus*. The reason for their dominance is not far-fetched. They are rich countries who spend large sums of their federal capital on research and development. Developing countries who wish to emulate their models will need higher fund allocations and attract higher numbers of skilled researchers in order to be able to compete.

## REFERENCES

Abrizah, A. and Wee, M.C. 2011. Malaysia's computer science research productivity based on publications in the Web of Science 2000-2010. *Malaysian Journal of Library & Information Science*, Vol.16, no.1: 109-124.

Al-Qallaf, C. 2003. Citation patterns in the *Kuwaiti Journal of Medical Principles and Practice*: The first 12 years, 1989-2000. *Scientometrics,* Vol.56, no.3: 369-382.

Anon. 1999. Outbreak of Hendra-like virus – Malaysia and Singapore 1998-1999. *Morb. Mort. Weekly Report*, Vol.48, no. 13: 265-269.

Bradford, S.C. 1948. *Documentation*. London: Crosby Lockwood.

Chiu, W.T. and Ho, Y.S. 2005. Bibliometric analysis of homeopathy research during the period of 1991 to 2003. *Scientometrics*, Vol.63, no.1*:* 3–23.

Chong, H.T., Seaton, B.T., Broder, C.C., Middleton, D. and Wang, L.F. 2006. *Hendra* and *Nipah viruses*: different and dangerous. *Nat. Rev. Microbiol*, Vol.4:23-35.

Chong, H.T., Suhailah, A. and Tan, C.T. 2009. *Nipah virus* and bats. *Neurology Asia,* Vol.14:73-76.

Chua, K.B., Bellini, W.J., Rota, P.A., Harcourt, B.H., Tamin, A., Lam, S.K., Ksiazek, T.G., Rollin, P.E., Zaki, S.R., Shieh, W.J., Goldsmith, C.S., Gubler, D.J., Roehrig, J.T., Eaton, B.T., Gould, A.R., Olson, J., Field, H.E., Daniels, P.W., Ling, A.E., Peters, C.J., Anderson, L.J. and Mahy, B.W.J. 2000. *Nipah virus*: a recently emergent deadly paramyxovirus. *Science***,** Vol.288**:** 1432-1435.

Chung, K.H. and Cox, R.A.K. 1990. Patterns of productivity in the finance literature: A study of the bibliometric Distributions, *The Journal of Finance,* Vol.45, no.1: 301-309.

Eaton, B.T. and Broder, C.C. 2006. *Hendra* and *Nipah Virus*: different and dangerous. *Nat. Rev. Microbiol*, Vol.4:201-207.

Glanzel, W. 2003, *Bibliometrics as a research field: A course on theory and application of bibliometric indicators*. Course Handouts.

Khean, J.G., Chong, T.T., Chew, N.K., Tan, P.S.K., Kamarulzaman, Sarji, S.A., Wong, K.T., Abdullah, B.J.J., Chua, K.B.C. and Lam, S.K. 2000. Clinical features of *Nipah virus encephalitis* among pig farmers in Malaysia, *New England Journal of Medicine*, Vol.342, no.17: 1229-1235.

Kugler, M. 2004. *Nipah Virus*: Emerging Infectious Disease. Rare Diseases About.com. Available at: http://rarediseases.about.com/od/rarediseasesn/a/091104.htm.

Lam, S.K. and Chua, K.B. 2002, Nipah virus encephalitis outbreak in Malaysia. *Clinical Infectious Diseases*, Vol.34, Suppl 2:S48-51.

Lotka, A.J. 1926. The frequency distribution of scientific productivity. *Journal of the Washington Academy of Sciences*, Vol.16, no.2: 317-324.

Paton, N.I., Leo, Y.S. and Zaki, S.R. 1999. Outbreak of *nipah-virus* infection among abattoir workers in Singapore, *Lancet*, No.354: 1253-1256.

Porotto, M., Rockx, B., Yokoyama, C.C., Talekar, A., Ilaria, D. Palermo, L.M., Liu, J., Cortese, R., Lu, Min, Feldmann, H., Pessi, A. and Moscona, A. 2010. Inhibition of *Nipah Virus* infection in vivo: targeting an early stage of paramyxovirus fusion activation during viral entry, *Plos Pathogens*, Vol.6, no.10: 28.

Ling, A. 1999. Lessons to be learnt from the "*Nipah Virus*" outbreak in Singapore. *Singapore Medical Journal*, Vol.40: 331-332.

Luby, S.P., Rahman, M., Hossain, M.J., Blum, L.S., Husain, M.M., Gurley, E., Khan, R., Ahmed, B.N., Rahman, S., Nahar, N., Kenah, E., Comer, J.A. and Ksiazek T.G. 2006. Foodbourne transmission of *Nipah virus*, Bangladesh. *Emerging Infectious Disease*, Vol.12: 1888-1894.

Melin, G. 2000. Pragmatism and Self-organization: Research Collaboration on the Individual Level, *Research Policy,* Vol.29, no.1: 31–40.

Nwagwu, W.E. 2007. Patterns of authorship in the biomedical literature of Nigeria. *Libres: Library and Information Science Research Electronic Journal*, Vol.17, no.1:1-28. Available at: http://libres.curtin.edu.au/libres17n1/NwagwuPatterns_Final_rev.pdf.

Reynes, J.M., Conner, D., Ong, S, Faure, C., Semg, V. and Molia, S. 2005. *Nipah virus* in Lyle's flying foxes, Cambodia. *Emerging Infectious Disease.* Vol. 7. Available at: http://www.cdc.gov/ncidod/EID/vol11no07/04-1350.htm.

Sanni, S.A. and Zainab, A.N. 2010. Google scholar as a source for citation and impact analysis for a non-ISI indexed medical journal. *Malaysian Journal of Library & Information Science*, Vol.16, no.3: 35-51.

Vermeulen, V., Parker, J.N. and Pender, B. 2010. Big, small or mezzo?, *EMBO Reports*, Vol.11, no.6: 1-4.

Wacharapluesadee, S., Lumlertdacha, B, Boongird, K, Wanghongsa, S, Chanhome, L, Rollin, P, Stockton, P., Rupprecht, C.E., Ksiazek, T.G. and Hemachudha, T. 2005. Bat *Nipah virus*, Thailand. *Emerging Infectious Disease,* Vol. 11: 1949-51.

Wong, K..,Shieh, W.J., Zaki, S.R. and Tan, C.F. 2002. *Nipah Virus* infection: an emerging paramyxoviral zoonosis, *Springer Seminar in Immunopathology*, Vol.24, no.2: 215-228.

Zainal, H, Z. and Zainab, A.N. 2011. Biomedical and health sciences research publication productivity from Malaysia. *Health Information and Libraries Journal*, Vol. 28, no.3: 216-224.

.